\documentstyle[11pt,paspconf,epsf]{article}

\begin{document}

\title{Observational Evidence for a Bar in the Milky Way}
\author{Konrad Kuijken}
\affil{Kapteyn Institute, PO Box 800, 9700 AV Groningen, Netherlands}

\begin{abstract}
Evidence from a variety of sources points towards the existence of a
bar in the central few kpc of the Galaxy.  The measurements roughly
agree on the direction of the bar major axis, but other parameters
(axis ratio, size, pattern speed) are still poorly
determined. Current dynamical models are limited by the quality of
hydro simulations, the degeneracy of stellar orbit models,
stellar-kinematic data and the significant lopsidedness of the central
kpc. Microlensing promises new constraints on the mass distribution in
the bulge/bar region.
\end{abstract}

\section{Introduction}
Our vantage point inside the Milky Way disk offers us the possibility
of a unique insight into the structure of at least our galaxy; however
this same location causes many unique problems. It is undoubtedly
useful to compare the Milky Way with other, more distant spiral
galaxies, but this rarely happens on an equal footing: sometimes the
Milky Way serves as a local well-understood calibrator for the
external galaxies, other times it is the other galaxies which provide
inspiration and guidance necessary for us to be able to interpret the
observations of our own Galaxy. Though the study of the dynamics of
the inner regions of the Milky Way may be said to be still in the
`borrowing from other galaxies' phase, the many data being gathered
make it likely that eventually we will be able to `give' as well.

This review covers the mounting evidence that the center of the Galaxy
harbours a bar with a size of a few kpc. Early evidence for a bar
(\S2) was championed mostly by de Vaucouleurs, but received little
following until interest was rekindled about five years ago by results
from near-infrared surveys. Since then, many detections of a barred
distortion, broadly consistent with each other (at least as regards
direction of the bar major axis) have appeared. They fall into two
broad categories, those based on photometric data (surface photometry
and star counts, \S3) and those based on kinematics of stars and gas
(\S4). More recently, the gravitational microlensing searches in the
direction of the Galaxy bulge have turned up many more events than had
been originally expected based on a simple axisymmetric model for the
Milky Way. A bar may significantly enhance these expected rates, and
may well be required to explain the microlensing data (\S5). This
review ends with a wish list of some observations which might help
constrain the bar parameters in the future (\S6).

\begin{table}
\caption{Properties of the Galaxy with a bearing on its Hubble type,
according to de Vaucouleurs (1970). Each property/morphological type
pair at stage Sbc is
assigned a score of $+1$ (good agreement), 0 (indifferent), or $-1$
(conflict).}
\label{tabdv}
\begin{centering}
\begin{tabular}{lccccccc}
Criterion & (a) & (b) & (c) & (d) & (e) & (f) & \multicolumn{1}{c}{Sum} \\
\tableline
A(s)   & $-$ & $-$ & $-$ & $-$ &   & $-$ & $-5$ \\
AB(s)  & $-$ & $-$ & $-$ & $-$ & + & $-$ & $-4$ \\
B(s)   & $-$ & $-$ & $-$ &     &   & $-$ & $-4$ \\
B(rs)  &     & +   &  +  &     &   &  +  &   +3 \\
B(r)   &     & +   &     &  +  &   &     &   +2 \\
AB(r)  & +   & +   &     &     &   &     &   +2 \\
A(r)   &     &     &     & $-$ &   & $-$ & $-2$ \\
A(rs)  &     &     &  +  & $-$ & + & $-$ &    0 \\
AB(rs) & +   & +   &  +  &     &   &  +  &   +4 \\
\tableline\tableline
\end{tabular}
\end{centering}
\tablenotetext{a}{High spiral arm multiplicity}
\tablenotetext{b}{Inner ring diameter of 6kpc}
\tablenotetext{c}{Broken ring structure}
\tablenotetext{d}{Radio structure of the nucleus}
\tablenotetext{e}{Yerkes spectral type}
\tablenotetext{f}{Non-circular {\sc Hi} motions}
\end{table}

\section{Early Evidence}

De Vaucouleurs (1964, 1970) early on suggested that the Milky Way was
in fact a barred spiral. His argument relied on comparing many
morphological features of the Milky Way with other spirals of
different revised Hubble types, the crucial step in this analysis
being to associate the {\sc Hi} feature known as the 3-kpc arm (also
interpreted in terms of a bar by Kerr 1967) as part of a broken {\sc
Hi} ring.  For a list of properties, he gave each of the subtypes
(r,rs,s) and (A,AB,B) a score reflecting their goodness of fit for the
Milky Way, and obtained an overall score by a simple unweighted sum
(see Table~1).  The best fit was the completely mixed type SAB(rs)bc:
a galaxy with fairly weak rings and a bar with not-quite grand design
spiral structure. It is interesting, and reassuring, that this
conclusion does not appear to be dominated by a single column in
Table~\ref{tabdv} (though three of the six diagnostics pertain to the
3-kpc expanding arm). There thus appeared to be a good case for taking
this suggestion seriously.

However, much of the effort in understanding Galactic structure in
subsequent years was focused on the problem of maintaining spiral
density waves, and the idea that the Milky Way had a bar fell out of
fashion.  Only in the last five years has it returned into favour.

\section{Photometric Evidence}

Because the Galaxy is virtually edge-on, we cannot see a bar
directly. However, unless the bar happens to be aligned along or
perpendicular to the
sun-Galactic center line, a bar will create systematic
differences between points at equal and opposite longitudes: if the
major axis of the bar lies in the first quadrant, for instance,
(longitude $0<\ell<90^\circ$) then objects in that quadrant will on
average be closer to the sun than those in the fourth quadrant. Such
effects can show up both in surface brightness and in star counts.

\subsection{Surface photometry}

Blitz \& Spergel (1991), in their search for non-axisymmetric
structure in the Milky Way, analysed the near-IR data of Matsumoto et
al.\ (1982) and showed that there were indeed systematic differences
between positive and negative longitudes near the Galactic center.
They showed that these could be understood as a perspective view of a
bar: the near side (in the first quadrant) would appear more
vertically extended on the sky than the far side, and the surface
brightness of the near side should be greater, both as observed.
Furthermore, in the innermost regions the asymmetry in surface
brightness is actually reversed, and this feature too is reproduced in
the data.

\begin{figure}
\plotone{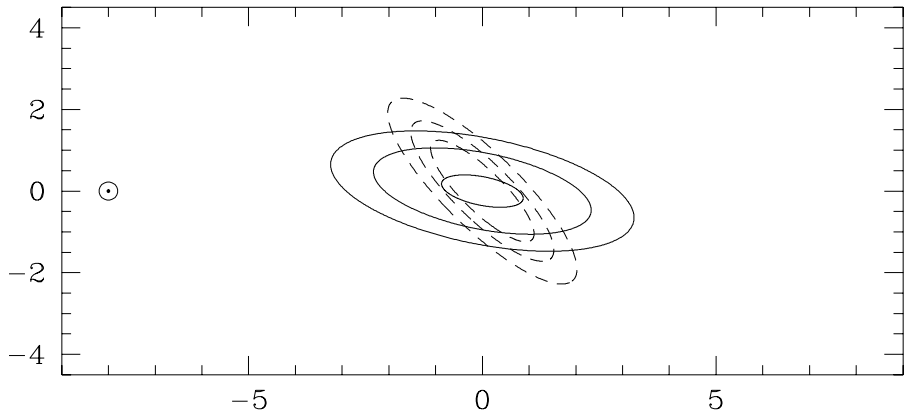}
\caption{The bar models `G2' (solid lines) and 'E3' (dashed)
of Dwek et al. (1995) projected onto the Galactic plane. Contours are
spaced at factors of 3 in mid-plane density; the outermost contour
corresponds to $3\times10^6L_\odot/\rm kpc^{-3}$.  The position of the
sun ($R_0=8$kpc) is indicated, and galactic longitude increases
counterclockwise from the right.}
\end{figure}
Superior data from the DIRBE experiment on the COBE satellite have
confirmed and sharpened this result. After dereddening the near-IR
data, Dwek et al. (1995) derive a best-fit model `G2' for the
emissivity of the bar of the form
\begin{equation}
\rho(x,y,z)\propto e^{-s^2/2}
\end{equation}
where (assuming the sun is 8kpc from the Galactic center)
\begin{equation}
s^4=\left[\left(x\over1490\rm
pc\right)^2+\left(y\over580\rm pc\right)^2\right]
+\left(z\over400\rm pc\right)^4
\end{equation}
and $(x,y,z)$ are Galactic coordinates rotated by $13.4^\circ$ about
the Galactic minor ($z$-) axis. This functional form implies an
ellipsoidal shape for the bar projected onto the galactic plane, but
makes boxy bulge isophotes as seen from earth, as observed. (To some
at this conference, the boxy isophotes are already strong evidence for
a bar!). The axis ratios are 2.6:1:0.7.  There are still uncertainties
in this deprojection, which was derived as a least-square fit to a
fairly restricted set of models---it will be hard to do better given
the usual problems associated with recovering a three-dimensional
emissivity from a two-dimensional surface brightness map. For
instance, model `E3', a triaxial version of Kent's (1992) modified
Bessel function model fits almost as well as G2, but with major axis
position angle $40^\circ$ (this model does have the unsatisfactory
feature that the z:y axis ratio is greater than 1).  Further
uncertainty arises from the correction for the disk contribution to
the surface brightness. A sketch of both bar models is shown in
Figure~1.

\subsection{Star counts}

Counts of individual objects also reveal left-right asymmetries of the
Galactic center region. All such data sets show an effect in the same
direction: objects at positive longitudes appear brighter and
therefore are presumably closer. These data sets include SiO masers
(Nakada et al. 1991, Izumiura et al.\ 1994), IRAS AGB stars (Weinberg
1992), IRAS Miras (Whitelock \& Catchpole 1992), the OGLE red clump
stars (Stanek et al. 1994), and OH/IR stars (Sevenster, this volume).
Furthermore, the globular clusters also appear to show a bar-like
distortion (Blitz 1993). Typical magnitude offsets are 0.2--0.5, the
best-measured one being that of the OGLE group ($0.37\pm0.03$).  For
comparison, the Dwek et al. (1995) G2 model would allow at most a 0.2
magnitude offset between brightness of objects at positive and
negative longtudes within $6^\circ$ of the galactic center (Figure~2).
\begin{figure}
\plotone{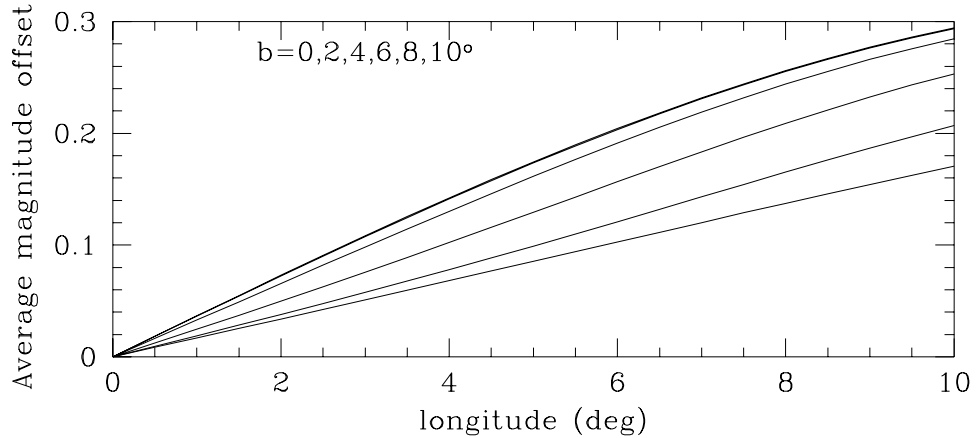}
\caption{The average magnitude offset between objects at galactic coordinates
$(\ell,b)$ and $(-\ell,b)$ in the Dwek et al. bar model. Different
curves correspond to different latitudes: (top to bottom)
$b=0,2,4,6,8,10^\circ$.}
\end{figure}

While all these studies agree on the sign of the asymmetry, the
agreement mostly ends there: the magnitude offset between positive and
negative $\ell$ varies considerably from survey to survey (though part
of the effect may be due to small number statistics in some of these
data sets, and different depths of the different samples). Also, the
results of the bulge surface photometry imply a very much smaller bar
than the IRAS AGB counts of Weinberg: the former extends out to
longitudes of about $10^\circ$, and the latter out to about
$40^\circ$. It is therefore quite possible that the Milky Way in fact
contains {\em several} bars.

It is interesting to note that, had we not known about bars in other
galaxies, we might not have chosen to interpret these left-right
asymmetries in such terms. On the sky, the asymmetries suggest a
lop-sided ($m=1$) distortion instead, and it might have seemed
far-fetched to attribute these to perspective effects of an inherently
$m=2$ bar.

\section{Kinematic Evidence}

Bars also show up as kinematic distortions of the velocity fields of
stars and gas, since the closed orbits in a pattern-rotating barred
potential are no longer circular, but rather elongated along or
perpendicular to the bar. Resonances (chiefly inner and outer Lindblad
and corotation) affect the orbit structure profoundly.  Some closed
orbits are self- or mutually intersecting, making them unsuitable as
gas orbits and consequently generating gaps and shocks in the gas
distribution (see the review by Athanassoula in this volume), and the
distribution of stellar orbits follows similar behaviour. Unlike
photometric signatures (except perspective effects discussed above),
these kinematic effects of a bar are also visible in edge-on systems,
and so provide a means of detecting bars in such galaxies.

Both the kinematics of gas and stars could reveal evidence for a bar
in the Milky Way, but each have their problems when it comes to
quantifying bar parameters. Gas, because it tends to dissipate down to
the closed, non-intersecting orbits, delineates the orbit structure
and hence the potential most clearly, but the most striking features
are associated with the resonances where hydro-dynamic effects are a
major factor. It is still very difficult to model all the relevant
processes at these locations well.  Stellar orbits, on the other hand,
are dominated by gravitational forces, but because of the absence of
dissipation the accessible orbits are much more varied. It is still an
unsolved problem to derive the gravitational potential from observed
radial velocity distributions in stationary elliptical galaxies, and
the barred galaxy problem, which also involves unknown figure
rotation, is even more complicated. Therefore, though it is possible
to rule out axisymmetric models on the basis of kinematic data,
producing a unique bar model is a considerably harder problem.

\subsection{Gas kinematics}

The distribution of gas within $5^\circ$ of the Galactic center is
complicated. Significant features for our purposes are:
\begin{itemize}
\item Large forbidden CO and {\sc Hi} velocities
\item A fast outwards decline in {\sc Hi} tangent point velocities
\item A {\em very} lopsided CO distribution
\item A dramatic change in the CO kinematics near longitudes $+1.7$
and $-1^\circ$.
\item A tilted (by $\sim7^\circ$ projected onto the sky) {\sc Hi}
(and maybe CO) plane.
\end{itemize}

The CO velocity structure, from the Bell Labs survey (Bally et al.\
1987, 1988) is shown in Figure~3.
\begin{figure}
\plotone{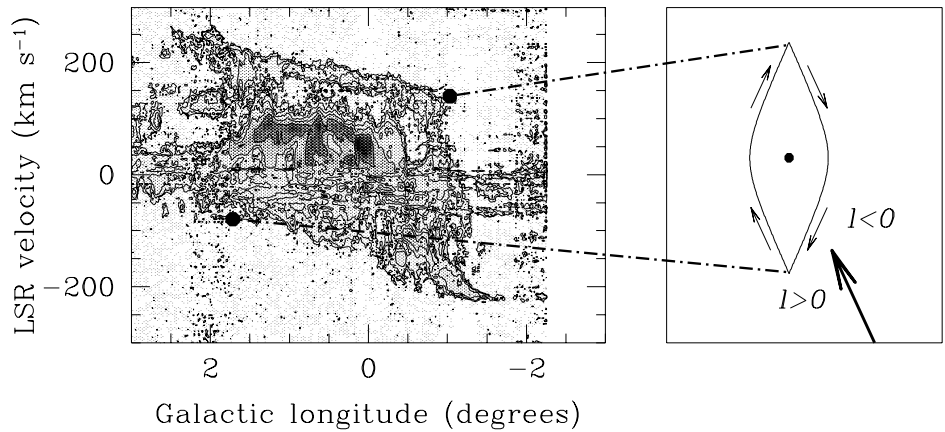}
\caption{Left: the distribution of CO emission in longitude and radial
velocity, from the Bell Labs.\ survey. Emission at $b=-3'$ is shown.
Contours are drawn at brightness temperatures of 1,2,4,8 and 16
K. Note the parallelogram-shaped envelope of the emission. Right: a
model for the parallelogram, from Binney et al.\ (1992). The cusped
orbit is viewed from the direction of the thick arrow, and gas streams
around the orbit as indicated.}
\end{figure}

The forbidden velocities (negative velocities in the first quadrant,
positive ones in the fourth) imply non-axisymmetry, assuming the gas
to be a dynamically cold tracer of the potential. However, per se they
say little about the nature of this deviation from circularity: in
particular, local expanding features in the gas may cause features
similar to those observed (e.g. Oort 1977, Uchida et al. 1994).

So far, no coherent {\em dynamical} model has been formulated which
addresses all observed features (but see Weinberg's paper in this
volume). However, several analyses have focused on subsets of these
observations. All these investigations have centered on single-bar
models, though reality may well be more complex.

Liszt \& Burton (1980) have modelled the kinematics of the {\sc Hi} in
the central few kpc as a tilted, elliptically streaming disk (an
earlier fit as an expanding disk was equally succesful, if less
plausible). Their model succesfully fits the observed distribution of
{\sc Hi} in position on the sky and radial velocity, though it offers
no dynamical origin for this disk. If the ellipticity is caused by a
bar, then the least satisfactory aspect of this model is the absence
of pattern rotation. It seems plausible, though, that a pattern speed
could fairly simply be accomodated in such a kinematic model.

Mulder \& Liem (1986) attempted to construct a global model for the
{\sc Hi}. Using non-selfgravitating hydrodynamical simulations
(pioneered in this context by Sanders \& Huntley 1976), they showed
that a multitude of kinematic features in the Galactic {\sc Hi} could
be explained with a simple model in which a gas disk is forced into a
quasi-steady flow by a simple model for a weak, rotating bar. In
particular, the 3kpc arm could be identified with shocked material
near the inner Linblad resonance, while the sun's position near
corotation (implying quite a slow pattern speed) explained the three
nearby spiral arms. Forbidden velocities in the central few degrees
could also be accounted for.  Their striking results, however, appear
to have received relatively little interest at the time.

Binney et al.\ (1992) concentrated on the distribution of the CO and
other molecular gas at $b=0$ (Fig.~3), and
interpreted it in terms of a dynamical model in which the gas is allowed
to move on closed, non-intersecting orbits only. No attempt was made
to address the tilt of the inner plane. They identify the striking
parallellogram shape of the orbit with the smallest orbit outside the
inner Lindblad resonance which does not intersect itself---gas further
in will strongly dissipate kinetic energy and end up in inside the ILR
on an `$x_2$ disk'. The parallelogram orbit is cusped, and seen from a
fairly narrow range of angles, its projection into the
longitude-velocity plane takes on the observed shape. Because this
orbit is strongly affected by the resonances, the pattern speed of the
Binney et al.\ model is very well constrained, with corotation around
$2.4\pm0.5\,$kpc. Furthermore, the parallellogram projection of the
orbit only appears from viewing angles of the bar about $16^\circ$ off
end-on. The distribution of {\sc Hi} at larger radii is consistent
with the closed orbits outside corotation, as is the radial dependence
of the model bar density with that of the observed bulge light.

In spite of these successes, a worrying aspect of this model is the
left-right asymmetry of the parallelogram. The data show such an
effect in the sense expected from perspective, but it is much more
pronounced than expected from the model. Rigorous modelling of the
observed asymetry (the cusps of the orbit appear at longitudes
$\sim+1.7^\circ$ and $-1^\circ$) implies that the cusps of the
parallelogram orbit lie at radius $\sim R_0/5\simeq1.6$kpc. An orbit
of this size would have to be aligned within $6^\circ$ of the line of
sight, and would have to be very slender if we were indeed viewing it
down its sides. A more plausible, if perhaps less elegant, explanation
invokes some lopsidedness to the central kinematics, which spoils the
perspective of the bar orbit. Such a component may be needed anyway to
explain the rather large velocity difference of the gas deduced to lie
near the cusps: this gas should have zero velocity in the bar frame.

The dynamics of the tilt of the inner gas are a puzzle. It may have
consequences for the bar analysis: when Liszt \& Burton (1980)
restrict their model to the Galactic plane (rather than the tilted
inner disk plane), velocity crowding of the gas mimicks the observed
distribution of CO very nicely. It therefore appears that the
identification of the parallellogram with a specific CO orbit is not
unique, and a more detailed consideration of the CO distribution out
of the galactic plane is required. (Initial suggestions by Blitz \&
Spergel (1991) that the stellar emission is tilted in the same
direction as the gas were shown by the COBE data to have been an
artefact of extinction. The stellar distribution is consistent with
being aligned with the Galactic plane).

Weiner \& Sellwood (1995) have concentrated on fitting the kinematics
of the {\sc Hi}, particularly the sharp falloff of the tangent point
velocity, outside longitudes $4^\circ$. They use a hydrodynamic code
to model the steady-state behaviour of the gas. Their results appear
inconsistent with the findings of Binney et al.: they find that only
bars seen over $30^\circ$ off end-on can generate forbidden velocities
over a sufficiently large longitude range. Their best-fit model also
has a significantly larger corotation radius of 3.6kpc.

The differences between the various analyses of the gas kinematics
partly reflect differences between the kinematics of the different
tracers, possibly due to nongravitational effects, but to some extent
also is an indication of the collisionality of the interstellar
medium: it remains a gross simplification to model gas as perfect
tracers of the closed non-intersecting orbits in a smooth,
pattern-rotating potential. For instance, the lop-sided distribution
of the central gas distribution is possibly a transient feature (e.g.,
a fluctuation associated with the relatively small number of large
clumps in the central few 100pc, or the result of a dynamical
instability or interaction) whose amplitude raises concerns about
fitting equilibrium bisymmetric models to the dynamics. An
investigation by Jenkins \& Binney (1994) shows that stochastic
processes in the gas distribution will indeed cause lopsidedness, but
they have difficulty reproducing effects as dramatic as those
observed. Quite possibly, self-gravity or low-temperature cooling
(neither of which is included in their calculations) can make a
substantial difference here.

Future refinements of the analyses of the observed gas dynamics may
well be inspired on observations of the CO distributions in
other barred galaxies (see reviews by Turner and Kenney in this
volume), which may establish when molecular gas does and does not
trace the closed orbits allowed by the potential.

\subsection{Stellar kinematics}

Given the possible problems with the dynamics of the gas, in how far
can stellar dynamics help?

At the moment, the answer is, unfortunately, not very much. The large
velocity dispersions in the bulge region wash out signatures of
non-axisymmetry, which only large numbers of stars sampled at a range
of longitude or integrated-light velocity distributions (see Kuijken
\& Merrifield 1995 or Merrifield, this volume) can overcome.

Apart from the difficulty of getting sufficiently detailed
observations, there is also a theoretical bottleneck: we do not know
what the velocity distributions in realistic galactic bars actually
look like, because there are large families of possible combinations
of stellar orbits which can be combined to make the same bar. Whereas
gas modelling can be simplified by considering closed orbits, this
constraint is not available in the case of stars. Ideally, it should
be replaced by a further observational phase-space measurement:
distance down the line of sight and/or proper motions. In any case,
just about the simplest axisymmetric model that can be constructed for
the bulge, an oblate isotropic rotator, appears to fit all available
stellar-kinematic data (Kent 1992), including recently published M
giant samples (Blum et al. 1995). This good fit is not evidence
against a bar, but rather an illustration of the difficulty of
detecting a bar in the stellar kinematics of the bulge. The strongest
feature in radial velocity data that argues in favour of a bar is the
bimodality of the OH/IR stars: in addition to a hot `bulge'
population, the central degree or so contains quite a cold stellar
disk, whose kinematics are similar to those of the inner CO gas
(Lindqvist et al. 1992). It is possible that these stars were formed
from the gas that lives inside the inner Lindblad resonance (an `$x_2$
disk').

The most detailed model constructed for the Milky Way's stellar bar is
that of Zhao (this volume). Analysis of the Spaenhauer et al.\
(1991) sample of stars with proper motion in Baade's Window (Zhao,
Spergel \& Rich 1994) shows possible signatures of triaxiality (vertex
deviations of metal-weak and metal-poor stars incompatible with
axisymmetry), but since the sample is small the statistical
weight of this study is rather low. Similar analyses of larger samples
in different parts of the bulge currently offer the best hope of
understanding the bar dynamics from stellar kinematics.

Long-range perturbations of the stellar kinematics by the quadrupole
field of a bar may also be detectable. Perturbation formulae for the
stellar velocity field, as well as the velocity dispersions, in a
barred, pattern-rotating potential, have been derived by Kuijken \&
Tremaine (1991). Weinberg (1994) has shown that the resonances of a
bar with a decreasing pattern speed will trace out a characteristic
signature across a disk, and he finds some evidence in the kinematics of
old K giants for such a feature.

\subsection{The pattern speed from stellar kinematics}

A particularly important product of stellar kinematics might be the
measurement of the pattern speed $\Omega_p$ of the bar.  Such
measurements can be made in model-dependent ways by identifying
certain morphological features (typically of the gas) with resonances,
or less so via an integral constraint derived from the continuity
equation (Tremaine \& Weinberg 1984, TW).  The TW method involves
integration along a given line of the velocity component perpendicular
to it. It was originally formulated for application to moderately
inclined external barred galaxies, in which case it requires
measurement of mean radial velocity along a line parallel to the major
axis. In that form it is inapplicable to edge-on galaxies such as our
own, but two modifications are: the first involves integration of
heliocentric radial velocities around the galactic equator (Kuijken \&
Tremaine 1991) and the second integration of transverse velocities
down a single line of sight near the Galactic center. Neither variant
is currently practical, however: the second requires accurate
distances and proper motions at $b=0$, whereas the first would rely on
full longitude coverage in the densest parts of the galactic plane,
with complete radial velocity coverage. Nevertheless, future large
near-infrared surveys may one day allow these measurements to be made.

\section{Gravitational Microlensing Evidence}

Microlensing of stars by foreground objects which pass at very small
projected impact parameters has recently developed from
an elegant curiosity to a new tool in galactic structure research. As
shown by Refsdal (1964), if a foreground object of mass $m$ at
distance $x$ from us passes within a radius
\begin{equation}
R_E(x)=2\sqrt{Gmx(1-x/L)}/c\propto m^{1/2}
\end{equation}
from the line of sight to a source at distance $L>x$, the
source will be magnified (`microlensed') by a factor $>1.34$. Since in
general the lens will move with respect to the line of sight, the
brightening will only last for a certain time, typically of the order of
1-100 days (depending on the lens mass). The average number of lenses
in the `microlensing tube' $R(x)<R_E(x)$ is called the optical depth
$\tau$, and depends on the number density $\nu$ of lenses along the
line of sight:
\begin{equation}
\tau=\int_0^L\nu(x)\pi R_E(x)^2dx\propto m^0\overline\rho
\end{equation}
where $\overline\rho$ is a mean mass density in lenses. $\tau$
therefore depends only on the mass density in lenses, not on the
masses of individual lenses (but the detectability does depend on $m$
via the timescale of typical events).

Whereas the detection rate for microlensing towards the Magellanic
clouds may be disappointingly low (Alcock et al.\ 1995a), the `control
experiments' towards the Galactic bulge have surprisingly turned up
many more events than had initially been expected: the optical depth
to the average bulge star is about $3\times10^{-6}$ (Udalski et al.\
1994, Alcock et al.\ 1995b). Along the line of sight towards Baade's
Window, a double exponential disk can produce at most
$\tau_D<1.2\times10^{-6}$, with a more likely number being less than
half that (Fig.~4).
\begin{figure}
\plotone{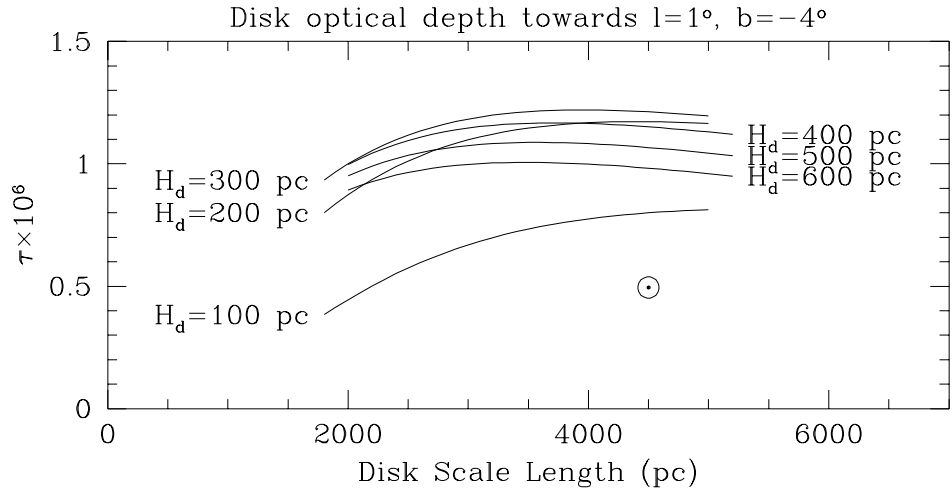}
\caption{The optical depth of bulge stars to microlensing by a double
exponential disk $\rho\propto \exp(-R/a-z/H_d)$, constrained to produce
a maximum rotation speed below 180km/s. The sun symbol shows the
optical depth due to a less than maximal disk, consistent with the
measurements of Kuijken \& Gilmore (1991).}
\end{figure}

The early calculations were based on axysymmetric models, and on the
assumption that the bulge stars were only lensed by foreground disk
and halo objects. It was later realized (Kiraga \& Paczinski 1994)
that bulge stars are so common that lensing of a far-side bulge star
by one on the near side contributes a significant signal
($\tau_B\sim0.7\times10^{-6}$ if one uses the Kent model for the
bulge). This signal is enhanced further if the bulge is extended along
our line of sight, for then the near-side stars are in a fatter part
of the microlensing tubes for lensing of the far-side bulge stars.
The effect can be as much as a factor of two if the bulge has the axis
ratio of the Dwek et al.\ (1995) model, raising the optical depth to
bulge sources to around $1.2\times10^{-6}$. The numbers are still a
little low, and larger numbers of microlensing events will have to be
analysed before it is clear whether there still is a problem or not.

Further constraints on bulge-bulge lensing can be derived by searching
for a systematic offset between the (unmagnified) brightnesses of
lensed stars with the general population. If far-side bulge sources
are systematically lensed more often than near-side ones, the lensed
sources should be systematically fainter. The magnitude of the offset
can be used to constrain the axis ratio and orientation of the bar
(Stanek 1995, and this volume).

\section{Conclusions and Wishlist of Further Observations}

It is clear that a variety of lines of evidence point towards the
existence of non-axisymmetric structure in the central few kpc of the
Milky Way. Equally impressive is the lack of evidence to the contrary!
While the precise details have not yet been characterised, rapid
progress is being made, partly driven by the need to understand the
new microlensing data.  Major stumbling blocks at the moment are the
difficulty of realistically simulating hydro\-dynamical processes.

In conclusion, it seems useful to compile a list of observations which
may help pin down the nature and parameters of the bar. These
include:
\begin{itemize}
\item {\em To see the bar in stellar kinematics.} Proper motions of
samples of stars throughout the bulge will greatly help define the
orbit structure, and hence the gravitational potential and pattern
speed of the bar.
\item{\em An optical depth map of the bulge region.} As shown by Kiraga
(1994), such a map provides an entirely separate constraint on the
mass distribution in the central regions.
\item{\em evolutionary history of the bulge as traced by stellar
abundances and their ratios.} Such data can be used to constrain the
star formation history of the bulge/bar, and combination with
kinematic data ultimately will allow the evolutionary relation between
the bar, bulge (if indeed they are separate) and disk to be addressed.
\item{\em Self-gravitating hydrodynamic simulations of gas flow in
barred potentials} will help address issues related to central
lopsidedness, stability and possible tilts.
\end{itemize}

\end{document}